\documentclass[aps,twocolumn,superscriptaddress,amsmath,amssymb,floatfix,noshowpacs]{revtex4}
\usepackage{graphicx,hyperref}
\usepackage{color}
\usepackage{soul}
\usepackage{amsmath}
\usepackage{tabularx}
\usepackage{array}
\usepackage{xcolor}
\usepackage[toc,page]{appendix}

\usepackage{multirow}
\usepackage{enumitem}
\usepackage[export]{adjustbox}  
\usepackage{changepage}
\usepackage[capitalise]{cleveref}
\graphicspath{{fig/}}

\newcommand{\hide}[1]{}

\crefformat{equation}{(#2#1#3)}

\newcommand*{\SIsupress}[1]{}

\begin{document}
\title{\textbf{Coupling light to an atomic tweezer array in a cavity}}
\author{Yakov Solomons}
\affiliation{Department of Chemical \& Biological Physics, Weizmann Institute of Science, Rehovot 7610001, Israel}
\author{Inbar Shani}
\affiliation{Department of Physics of Complex Systems, Weizmann Institute of Science, Rehovot 7610001, Israel}
\author{Ofer Firstenberg}
\affiliation{Department of Physics of Complex Systems, Weizmann Institute of Science, Rehovot 7610001, Israel}
\author{Nir Davidson}
\affiliation{Department of Physics of Complex Systems, Weizmann Institute of Science, Rehovot 7610001, Israel}
\author{Ephraim Shahmoon}
\affiliation{Department of Chemical \& Biological Physics, Weizmann Institute of Science, Rehovot 7610001, Israel}
\date{\today}

\begin{abstract}
We consider the coupling of light, via an optical cavity, to two-dimensional atomic arrays whose lattice spacing exceeds the wavelength of the light. Such `superwavelength' spacing is typical of optical tweezer arrays. While subwavelength arrays exhibit strong atom-photon coupling, characterized by high optical reflectivity in free space, the coupling efficiency of superwavelength arrays is reduced due to collective scattering losses to high diffraction orders. We show that a moderate-finesse cavity overcomes these losses. 
As the scattering losses 
peak at certain discrete values of the lattice spacing, the spacing can be optimized to achieve efficient atom-photon coupling in the cavity. Our cavity-QED theory properly accounts for collective dipolar interactions mediated by the lossy, non-cavity-confined photon modes and for finite-size effects of both the array and the light field. 
These findings pave the way to harnessing the versatility of tweezer arrays for efficient atom-photon interfaces in applications of quantum computing, networking, and nonlinear optics. 

\end{abstract}
 \maketitle

Spatially ordered arrays of atoms trapped by optical tweezers have emerged as a notable platform for quantum science and technology \cite{Nogrette,Lester,Lee,Barredo1,Endres,Barredo2,Schymik,Browaeys,Kaufman,DeMille,Bernien,Levine1,Levine2,Madjarov}. 
Here, we discuss the promising potential of establishing an efficient quantum interface between such atomic arrays and light. The motivation is twofold. First, viewing tweezer arrays as stationary nodes of quantum information systems, their coupling to light is essential for quantum networking or for distributed quantum computing \cite{Kimble,Monroe,Reiserer,Northup}. Second, the nonlinearity and versatile tools developed for atomic arrays are useful for generating and studying quantum many-body states of light \cite{coop_Chang,Perczel,Rivka,Bettles,cirac}.

The coupling of atomic arrays to light is a collective phenomenon, established by induced dipole-dipole interactions between the atoms. As such, it depends on the positions of the atoms and, more specifically, on the ratio between the array lattice spacing $a$ and the wavelength of light $\lambda$. A subwavelength lattice $a<\lambda$, typical of optical lattices \cite{Rui,Bloch1,Bloch2}, exhibits highly efficient light-matter coupling characterized by the array's strong reflectivity \cite{Bettles_mirror,Facchinetti,EfiPRL,Manzoni,Henriet,Asenjo-Garcia,Grankin,Facchinetti2,Guimond,Efi1,Efi2,Parmee,Patti,ALJ1,POHs,Plankensteiner,MALZ,Wei,Moreno,Zhang,Srakaew}. Conversely, in a superwavelength lattice  $a>\lambda$, typical of tweezer arrays \cite{Rivka,Caldwell}, the atoms are coupled not only to the normal-incidence beam but also to higher diffraction orders \cite{EfiPRL,Our_paper}. This leads to scattering losses and hence to a drastic reduction of the reflectivity and the efficiency of quantum operations \cite{Our_paper}. 

\begin{figure}[h!]
    \centering	
		\includegraphics[width=0.99\columnwidth]{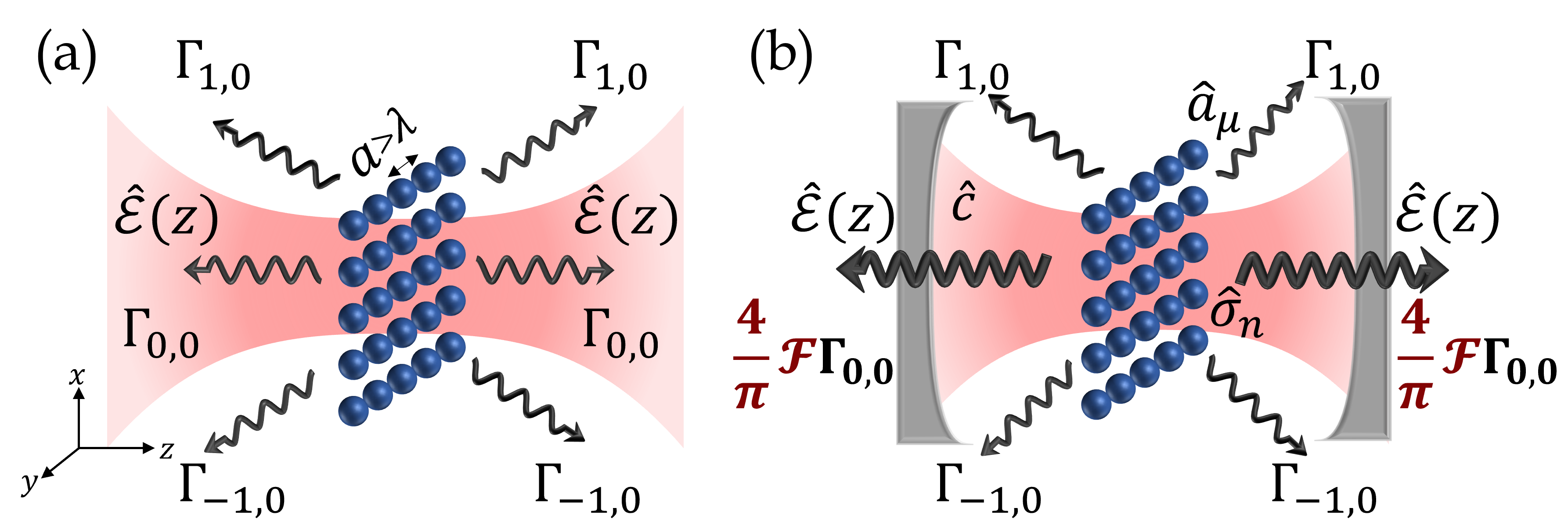}
		\caption{\small{
				(a) Coupling light to a square 2D array in free space: the array with lattice spacing $a$ is located on the $xy$ plane ($z=0$) and illuminated from both sides by a paraxial target mode $\mathcal{\hat{E}}(z)$, corresponding to the (0,0) diffraction order. In the superwavelength case $a>\lambda$, scattering also exists to higher diffraction orders $(m_x,m_y)$, 
    thus contributing to losses [Eq.~\cref{eq:diff}]. Here, the orders $(-1,0)$ and $(1,0)$ are marked, and $\Gamma_{0,0}\equiv\Gamma_0$ [Eq.~$\cref{Gamma0}$]. (b) A 2D array inside an optical cavity. The coupling to the (0,0) order is enhanced by the cavity finesse $\mathcal{F}$, while the coupling to all other diffraction orders is unaffected. 
		}}
		\label{2D_array}
\end{figure}

We propose to combat this reduction by using an optical cavity, enhancing the coupling to a desired ``target mode" beam. The idea is that while multiple reflections of the target mode from the cavity mirrors strengthen its coupling to the atoms, this is not the case for the light scattered to other diffraction orders. 
However, in order to properly analyze the coupling of tweezer arrays to light using a cavity, one has to go beyond the typical theory of cavity QED and include collective effects in the light scattered outside of the cavity. 

To this end, we study light-matter coupling of a two-dimensional (2D) superwavelength atomic array placed inside a cavity, while properly accounting for collective dipole-dipole interactions mediated by non-cavity-confined modes, as well as the finite sizes of the array and the cavity photon mode. We reveal two main features: (i) The desired coupling $\Gamma\propto \mathcal{F}$ to the target mode is enhanced by the cavity finesse $\mathcal{F}$. (ii) Scattering losses $\gamma_{\mathrm{loss}}$ are dominated by collective scattering to higher diffraction orders outside the cavity. Crucially, these losses depend strongly on the array lattice spacing, with large losses appearing at specific resonant values of the lattice spacing; these resonances are further accompanied by a non-trivial dependence on cavity-mode waist arising from competing effects. Accordingly, our analytical results, supported by numerical scattering calculations,
reveal the possibility of designing near-unity coupling efficiencies $r_0\equiv \Gamma/(\Gamma+\gamma_{\mathrm{loss}})$ by a proper choice of the lattice constant, even for moderate finnese. This opens the way for achieving  high fidelities of various quantum tasks performed by the tweezer-array quantum interface.\\

\emph{General formalism.---}
We consider a 2D tweezer array consisting of $N$ identical two-level atoms, forming a square lattice on the $xy$ plane with a superwavelength lattice spacing $a>\lambda$, $\lambda$ being the central wavelength of the incident light. In the following, we will first consider an array in free-space (\cref{2D_array}a) and then proceed to the case where it is placed inside a cavity (\cref{2D_array}b). In both cases, we will study how the array is coupled to a paraxial ``target mode" of interest, $\mathcal{\hat{E}}(z)=\frac{1}{\sqrt{2}}(\mathcal{\hat{E}}_+(z)+\mathcal{\hat{E}}_-(z))$, illuminating the array from both sides ($\pm$ for right- and left propagating beams, respectively) \cite{comment}. To this end, we follow the approach of Ref.~\cite{Our_paper} wherein many-atom, multimode problems are mapped to a 1D scattering problem, of a target mode $\mathcal{\hat{E}}$ with a transverse mode profile $u(\mathbf{r}_{\bot})$ [$\mathbf{r}_{\bot}=(x,y)$] which is coupled to a spatially matched collective-dipole $\hat{P}$,
\begin{equation}
\hat{P}=\frac{a}{\sqrt{\eta}}\sum_{n=1}^N u^*(\mathbf{r}^{\perp}_n)\hat{\sigma}_n,
\label{transform_P}
\end{equation}
where $\hat{\sigma}_n 
$ is the two-level lowering operator of atom $n$, and 
\begin{equation}
\eta=\int_{L_a^2}|u(\mathbf{r}_{\perp})|^2d\mathbf{r}_{\perp}\quad\xrightarrow[\mathrm{Gaussian}]{}\quad\eta=\mathrm{erf}^2(\frac{L_a}{\sqrt{2}w}),
\label{eta}
\end{equation}
is the fraction of the spatial mode that overlaps with the atomic array of size $L_a^2=a^2N$. For a typical case of a Gaussian target mode, $u(\mathbf{r}_{\perp})=\sqrt{\frac{2}{\pi w^2}}e^{-(\mathbf{r}_{\perp})^2/w^2}$ with waist $w\gg \lambda$, excellent overlap $\eta\rightarrow1$ exists for a sufficiently large array $L_a\gg w$. 

Within the general 1D description, the coupling between the target mode and the collective dipole in the low-excitation, linear regime (where $\hat{P}$ is bosonic, $[\hat{P},\hat{P}^\dagger]=1$) is captured by the Heisenberg equations \cite{Our_paper}
\begin{align}
 \label{EOM}
	\frac{d\hat{P}}{dt}&=-\left[\frac{\Gamma}{2}+\frac{\gamma_{\mathrm{loss}}}{2}+i\left(\Delta-\delta\right)\right]\hat{P}
	+i\sqrt{\Gamma}\mathcal{\hat{E}}_{0}(0)+\hat{F},
 \nonumber\\
 \mathcal{\hat{E}}(z)&=\mathcal{\hat{E}}_{0}(z)+i\sqrt{\Gamma}\hat{P}.
\end{align}
Here $\delta$ is the detuning between the central frequency of the target mode and the bare atomic transition, $\Gamma$ is the coupling to the target mode (whose input field is $\mathcal{\hat{E}}_{0}$),  $\gamma_{\mathrm{loss}}$ is the loss rate to undesired modes, $\Delta$ is a collective resonance shift, and $\hat{F}$ is the quantum Langevin noise corresponding to $\gamma_{\mathrm{loss}}$. 
The quality of the array-light interface in the 1D model is fully characterized by the cooperativity $C$, or equivalently, by the coupling efficiency $r_0$, defined as
\begin{equation}
C = \frac{\Gamma}{\gamma_{\mathrm{loss}}},\quad r_0 = \frac{C}{1+C}.
\end{equation}
The cooperativity $C$ is the branching ratio between the emission to the target mode and the loss, while $r_0$ is the fraction of energy absorbed (radiated) from (to) the target mode. In a two-sided system, where light can either be transmitted or reflected, $r_0$ is also equal to the on-resonance reflectivity. These parameters determine the fidelity of various quantum tasks, \textit{e.g.}, the quantum memory efficiency, or the maximal level of quantum correlations and entanglement achievable when adding non-linearity \cite{Our_paper}. We thus require $\Gamma\gg\gamma_{\mathrm{loss}}$ for an efficient light-matter coupling.\\

\emph{Array in free space.---}
In order to understand the sources of scattering losses of the array-light interface, it is instructive to consider first  
an array in free space. The mapping of the dynamical equations to those of Eqs.~\cref{EOM} is performed by assuming: (i) the array is large enough, $L_a\gg\lambda$, to take collective effects due to dipole-dipole interactions as in an infinite array, (ii) the target mode is paraxial enough, $w\gg\lambda$, to neglect spatial dispersion effects in the array response (corrections are discussed below). With these assumptions, we obtain Eqs.~\cref{EOM} with the parameters \cite{EfiPRL, Our_paper}
\begin{equation}
\begin{split}
&\Gamma^{(\mathrm{free})}=\eta\Gamma_0=\eta\frac{3}{4\pi}\frac{\lambda^2}{a^2}\gamma,\quad
\gamma_{\mathrm{loss}}^{(\mathrm{free})}=\gamma_\mathrm{s}+(1-\eta)\Gamma_{0}+\gamma_{\mathrm{diff}},
\label{Gamma0}
\end{split}
\end{equation}
and $C^{(\mathrm{free})}=\Gamma^{(\mathrm{free})}/\gamma_{\mathrm{loss}}^{(\mathrm{free})}$. Here $\Gamma_0$ represents the collective emission rate of a uniformly excited infinite array, and $\gamma$ is the free-space spontaneous emission rate of a single atom. \SIsupress{, and the collective shift $\Delta$ is given in \cite{SI}.} 

The three terms comprising $\gamma_{\mathrm{loss}}^{(\mathrm{free})}$ originate from three distinct loss mechanisms. The first two terms are associated with imperfections, with $\gamma_\mathrm{s}$ being the scattering loss due to non-collective processes such as weak disorder, and $(1-\eta)\Gamma_{0}$ representing the loss due to imperfect overlap of the target mode with the array (with $\gamma_s\ll\Gamma_0$ assumed negligible in the following). The third term $\gamma_{\mathrm{diff}}$ exists only in a superwavelength lattice ($a>\lambda$), and it describes the emission to higher diffraction orders, 
\begin{align}
\begin{split}
	\gamma_{\mathrm{diff}}&=\Gamma_0\sideset{}{'}\sum_{m_x,m_y}\frac{1-\frac{\lambda^2}{a^2}|(m_x,m_y)\cdot\textbf{e}_d|^2}{\sqrt{1-\frac{\lambda^2}{a^2}(m_x^2+m_y^2)}}\equiv\sideset{}{'}\sum_{m_x,m_y}\Gamma_{m_x,m_y}.
\end{split}
\label{eq:diff}
\end{align}
Here, the sum $\sum'$ is preformed over all propagating diffraction orders $(m_x,m_y)\neq (0,0)$ satisfying the condition ${|(m_x,m_y)|<a/\lambda}$. 

\begin{figure}[t]
	\begin{center}
        \includegraphics[width=\columnwidth]{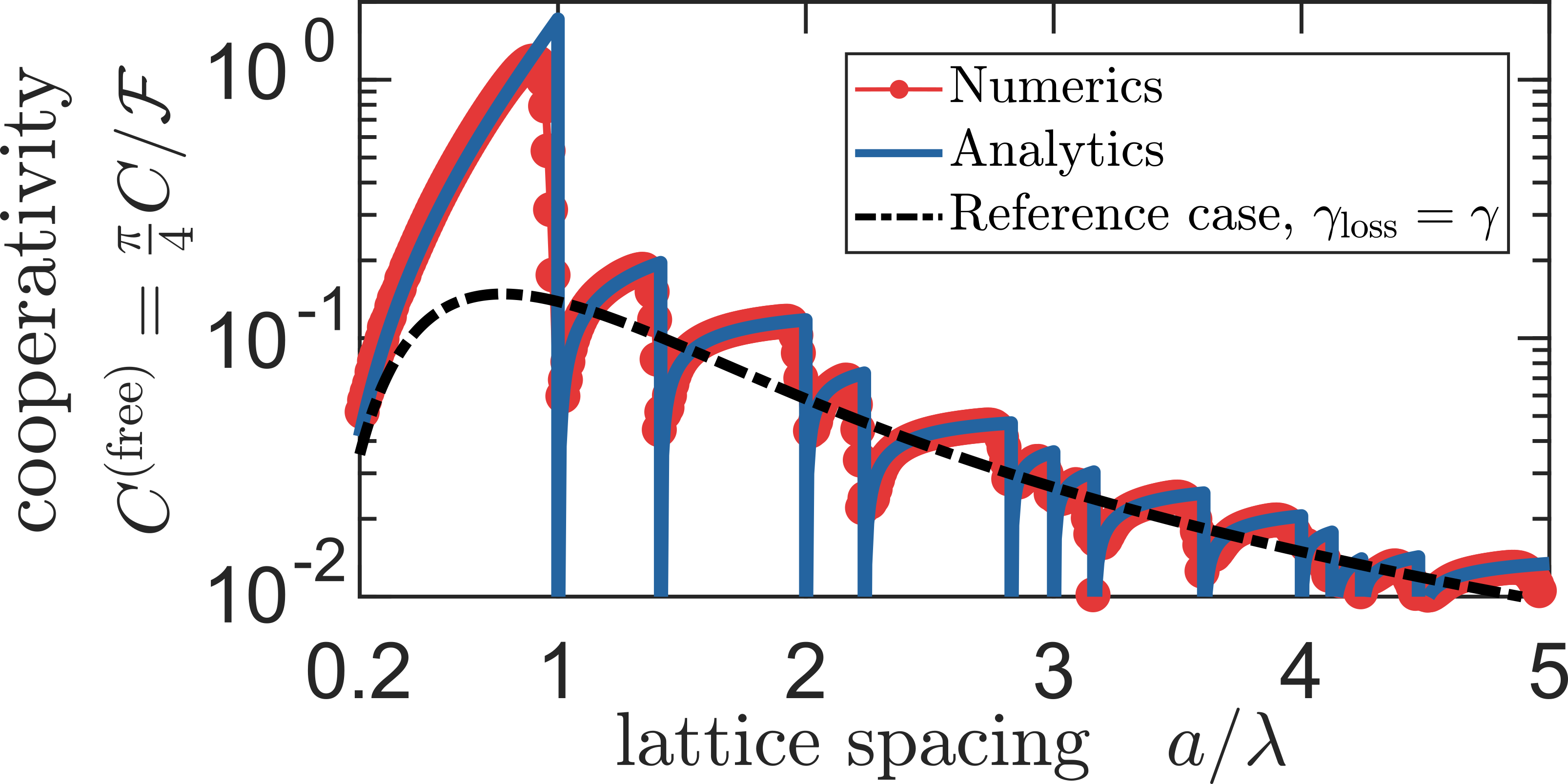}
		\caption{\small{
				The cooperativity of an atomic array in free space and in an optical cavity, $C^{(\mathrm{free})}=\frac{\pi}{4}C/\mathcal{F}$ [Eqs.~\cref{eq:gamma_cavity} and \cref{Gamma0}-\cref{eq:diff}].
    The numerical simulation (red circles) and analytical solution (solid blue) are calculated for an array of $N=20\times20$ atoms and a beam waist $w=15\lambda$. The cooperativity changes rapidly around the points $a/\lambda=\{1,\sqrt{2},2,\sqrt{5}...\}$, where new diffraction orders appear. The reference case (dashed black) describes results with individual-atom scattering losses $\gamma_{\mathrm{loss}}=\gamma$.
		}}
		\label{C_vs_a}
	\end{center}
\end{figure}

In an imperfect subwavelength array, enhancing the coupling $\Gamma$ to the target mode using a cavity is of course beneficial for increasing $C=\Gamma/\gamma_{\mathrm{loss}}$. However for a superwavelength array, it becomes practically \textit{essential} even in a perfect infinite system, due to the scattering  $\gamma_{\mathrm{diff}}>0$ to multiple diffraction orders which always exists and often dominates the loss, as seen in \cref{C_vs_a} and discussed further below.

\emph{Array inside an optical cavity.---} 
Motivated by the above conclusions, we proceed to consider an array inside a cavity. We assume that the target mode spatially matches the cavity mode so that its coupling to the array is enhanced, whereas the additional diffraction orders are all directed at angels which are not supported by the cavity mirrors and are thus scattered away, see \cref{2D_array}b. Since the scattering to the high diffraction orders is collective, the model must take into account dipole-dipole interactions mediated not only by the cavity modes but also by the non-confined 3D modes. To this end, we develop a modified cavity QED formalism that accounts for such collective effects, adopting the approach recently proposed in \cite{Efi4} to our case of a superwavelength array.   

We begin with the Hamiltonian of the full many-atom multimode system depicted in \cref{2D_array}b. \SIsupress{(see \cite{SI}).} The atoms $\hat{\sigma}_n$ are coupled to the cavity mode $\hat{c}$ via the Hamiltonian term $\hat{H}_{AC}=\hbar\sum_n [g_n\hat{c}\hat{\sigma}_n^\dagger+\mathrm{h.c}]$, with $g_n\propto u(\mathbf{r}^{\perp}_n)$. In turn, the cavity mode is coupled with a damping rate $\kappa$ to the 1D continuum representing the propagating target mode $\mathcal{\hat{E}}$ from both sides. In addition, the atoms are coupled to a photon reservoir comprised of the modes $\hat{a}_\mu$ which are not confined by the cavity. \SIsupress{As detailed in \cite{SI},} We first write Heisenberg-Langevin equations for the atomic and cavity variables by eliminating the non-confined modes $\hat{a}_\mu$ within a Markov approximation. For the equations of the atomic variables $\hat{\sigma}_n$, the non-confined modes contribute a collective interaction term between dipoles of different atoms, with the dipole-dipole kernel given by a partial photonic Green's function, $G_{\mathrm{nc}}$, which contains only the non-confined modes (nc). In general, finding the non-confined modes and $G_{\mathrm{nc}}$ is a system-specific, non-trivial task. However, a generic approximation can be made noting that: (i) the dispersive shift (given by $\mathrm{Re}[G_{\mathrm{nc}}]$) is dominated by near fields and is hence similar to free space, assuming the array is at least a wavelength away from the cavity mirrors; (ii) the emission to zeroth diffraction order is into the cavity, and hence must not be included in the losses due to the non-confined modes  (given by the $\mathrm{Im}[G_{\mathrm{nc}}]$). These considerations are accounted for by the following approximation \cite{Efi4,Efi3} \SIsupress{\cite{SI,Efi4,Efi3}}
\begin{equation}
{\rm Re}[G_{\mathrm{nc}}] \simeq {\rm Re}[G],
\quad\quad
{\rm Im}[G_{\mathrm{nc}}] \simeq {\rm Im}[G] - {\rm Im}[G_{\mathrm{c}}].
\label{Gnc}
\end{equation}
Here $G$ is the free-space Green's function of the photons, and $G_{\mathrm{c}}$ is the Green's function of the 1D continuum corresponding to the transverse cavity mode, which does not contain the modes corresponding to the additional diffraction orders that diverge aside. 

With this approximation and a subsequent adiabatic elimination of the fastly decaying cavity mode $\hat{c}$, \SIsupress{we show in \cite{SI} that} we arrive at the equations of the form \cref{EOM}, with the parameters
\begin{equation}
	\label{eq:gamma_cavity}
\Gamma=\frac{4}{\pi}\mathcal{F}\Gamma^{(\mathrm{free})},\quad\gamma_{\mathrm{loss}}=\gamma_{\mathrm{loss}}^{(\mathrm{free})},
\quad
C=\frac{\Gamma}{\gamma_{\mathrm{loss}}}=\frac{4}{\pi}\mathcal{F}C^{(\mathrm{free})}.
\end{equation}
Namely, compared to the free-space case, the coupling to the target mode obtains an enhancement factor proportional to the cavity finesse $\mathcal{F}$, while the scattering losses remain the same, since the higher diffraction orders are spanned by the non-confined modes described by $G_{\mathrm{nc}}$ in Eq.~\cref{Gnc}. 

The enhancement of the cooperativity $C$ by the finesse generalizes known results for non-collective behaviour in a cavity, \textit{e.g.}, a single atom or a dilute ensemble \cite{Hammerer,Lahad}, to the collective case of an array. 
Indeed, considering a reference case of an atomic ensemble with density $\rho$ and length $L_z$ placed in a cavity, and neglecting collective dipolar interactions via non-confined modes by assuming a sufficiently dilute ensemble, one obtains the individual-decay loss rate $\gamma_{\mathrm{loss}}^{(\mathrm{ref})}=\gamma$ and a coupling rate $\Gamma^{(\mathrm{ref})}$ of the form \cref{eq:gamma_cavity} but with the average transverse distance $a_{\mathrm{ref}}=\sqrt{\frac{1}{\rho L_z}}$ replacing the array lattice spacing $a$. \SIsupress{\cite{SI}.}

\emph{Diffraction resonances.---} 
With Eq.~\cref{eq:gamma_cavity} at hand [alongside Eqs.~\cref{Gamma0}-\cref{eq:diff}], we now examine the cooperativity $C$ as a function of the lattice spacing, for a given finite atom number $N$ and target-mode waist $w$. The result is plotted in \cref{C_vs_a}, where $C$ is scaled by $(4/\pi)\mathcal{F}$ to ease the comparison with the free-space result $C^{(\mathrm{free})}$. We validate our analytical results using numerical calculations of classical scattering in the simpler case of an array in free space illuminated by a Gaussian beam with the same waist. There, $C^{(\mathrm{free})}$ is obtained from the efficiency $r_0^{(\mathrm{free})}=C^{(\mathrm{free})}/(1+C^{(\mathrm{free})})$, which we recall to be given by the on-resonance reflectivity estimated from the scattered field as described in Refs.~\cite{EfiPRL,Our_paper}, finding excellent agreement with the analytical results [see Fig. \cref{C_vs_a}; sharper features in the analytical curve are due to the infinite-array approximation in Eq. \cref{eq:diff} for $\gamma_{\mathrm{diff}}$].

We observe that the cooperativity $C$ in \cref{C_vs_a} does not change smoothly with the lattice spacing $a$ but rather drops rapidly to zero at specific values. These values occur at $a/\lambda=\{1,\sqrt{2},2,\sqrt{5}...$\} where the diffraction losses in Eq.~\cref{eq:diff} exhibit divergences, corresponding to the emergence of new dissipation channels: diffraction orders that change from being evanescent (non-dissipative) to propagating (radiative). Enhanced reflectivity is obtained for lattice spacings $a$ just before new diffraction orders appear, corresponding to minima of $\gamma_\mathrm{diff}$ in Eq.~\cref{eq:diff}. 
This non-trivial behavior is in stark contrast to the reference case, the prediction of a typical cavity QED theory that neglects collective effects in the emission to non-confined modes. 
Our collective theory is therefore essential for designing efficient interfaces, particularly by choosing the optimal values of $a$ just before the resonances. 


\begin{figure}[t]
	\begin{center}
  		\includegraphics[width=1\columnwidth]{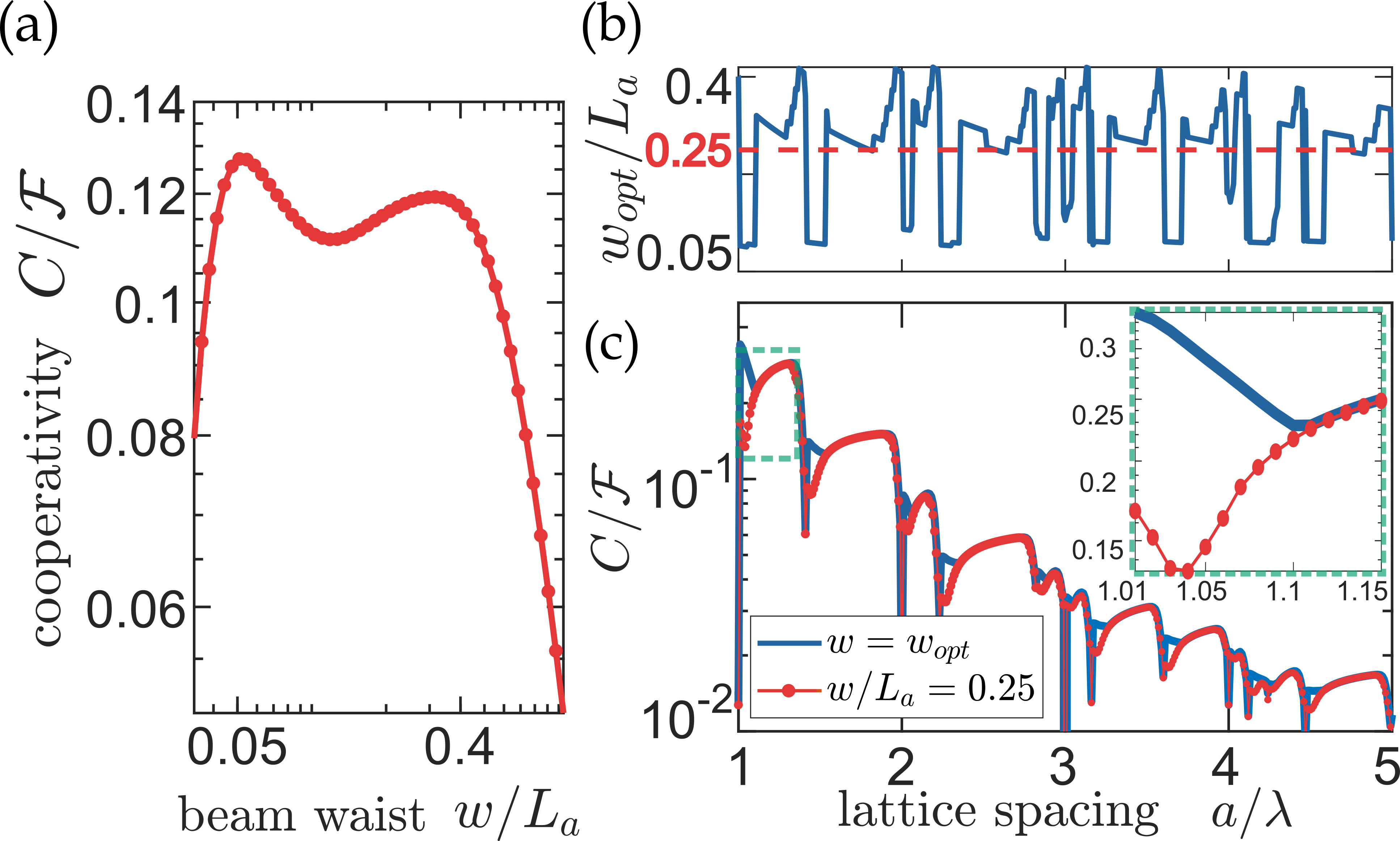}
		\caption{\small{ Finite size effects for an array of $N=20\times20$ atoms in a cavity. 
				(a) Cooperativity versus the beam waist $w$ close to the diffraction resonance $a/\lambda=\sqrt{2}+0.1$. (b) The optimal waist versus the lattice spacing. The optimal waist is close to the average value $w/L_a=0.25$, besides the jumps around the diffraction resonances. (c) The cooperativities for the optimal waists from (b), and for the average value $w/L_a=0.25$ become distinct closer to the diffraction resonances (see inset). The results are obtained via the numerical scattering calculations yielding $C^{(\mathrm{free})}=\frac{\pi}{4}C/\mathcal{F}$.
		}}
		\label{optimal_w}
	\end{center}
\end{figure}
\emph{Finite size effects.---} We now show how the diffraction resonances also lead to non-trivial behavior of the optimal waist. So far, the finite size of the array and beam was taken into account via the error function $\eta$. While this 
is an excellent approximation for a paraxial light-beam, it clearly favors small beams with respect to the array size. However, when optimizing the beam size, one has to take into account a competing \textit{dispersion effect}. Namely, as the beam waist narrows, non-zero transverse momentum $\textbf{k}_\perp$ components, which correspond to different collective dipoles of the array with different resonances $\Gamma_{\textbf{k}_\perp}/2+i\Delta_{\textbf{k}_\perp}$, become more significant, and the frequency of the beam cannot match all $\Delta_{\textbf{k}_\perp}$ simultaneously, resulting in reduced cooperativity \cite{Manzoni,EfiPRL}.

In the case of a superwavelength array, the physics is richer. An additional effect occurs around the diffraction resonances, which we refer to as the \textit{diffraction effect}. As discussed earlier, when the lattice spacing crosses a point of a new diffraction order, determined by Eq.~\cref{eq:diff}, the cooperativity drops rapidly to zero. However, for the non-zero transverse momentum components $\textbf{k}_\perp$, the diffraction orders appear at different points, since for them the wavevector $\frac{1}{a}(m_x,m_y)$ in \cref{eq:diff} changes to $\frac{1}{a}(m_x,m_y)+\textbf{k}_\perp/(2\pi)$. Therefore, a smaller waist with a larger spread in momentum space could reduce the rapid change in the cooperativity. As a result, the dependence of $C$ on the waist may exhibit two maxima. The first maximum results from the balance between the dispersion effect, which favors a larger beam waist, and the diffraction effect, which favors a small beam waist. The second maximum results from the balance between the dispersion effect and the spatial overlap mismatch between the beam and the array. Figure \ref{optimal_w}a shows the cooperativity close to the appearance of the second diffraction order $a/\lambda=\sqrt{2}+0.1$. Two maxima are seen 
providing two optimal regimes to work in.

Figure \ref{optimal_w}b shows the optimal waist as a function of the lattice spacing. The optimal waist, which is generally close to its average value $w/L=0.25$, exhibits jumps around the diffraction resonances, corresponding to the two maxima as demonstrated in \cref{optimal_w}a. Figure \ref{optimal_w}c shows the cooperativity for these numerically found optimal waists. We find that around the diffraction resonances, the exact optimal waist gives better results than its average value, while far from these points, they coincide.\\

\begin{figure}[t]
	\begin{center}
		\includegraphics[width=1\columnwidth]{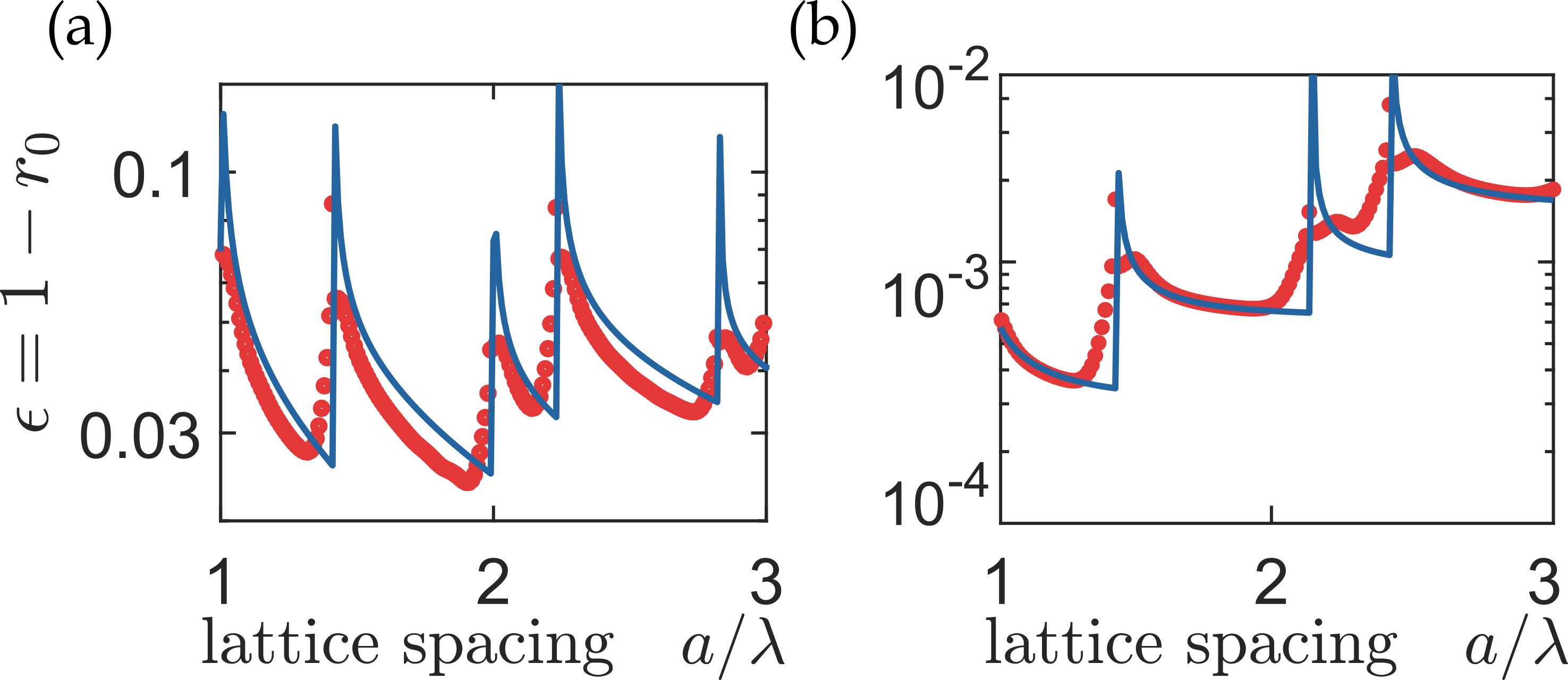}
		\caption{\small{
				The coupling inefficiency $\epsilon=1-r_0$  for an array of $N=10\times10$ atoms in a cavity (numerical calculations in red dots; analytical results in solid blue). (a) For a cavity finesse $\mathcal{F}=1000$ and beam waist $w/\lambda=25$. (b) For a high finesse $\mathcal{F}=10000$ and $w/\lambda=5$. At $a/\lambda\lesssim \sqrt{2}$, the inefficiency reduces to $\epsilon\sim3.5\times 10^{-4}$.
		}}
   		\label{realistic}
	\end{center}
\end{figure}

\emph{Realistic example.---}
The cavity enhancement is readily achievable with realistic parameters. Figure \ref{realistic}a shows the coupling inefficiency $\epsilon=1-r_0=\frac{1}{1+C}$ for an array of $N=10\times10$ atoms placed inside a cavity with finesse $\mathcal{F}=1000$ for a beam waist $w=25\lambda$ (for example, $w\approx 20~\mu$m for $^{87}$Rb atoms with $\lambda=0.78~\mu$m). It is seen that even for lattice spacing of about $a\approx1.9\lambda$, the inefficiency is smaller than $2\%$. Interestingly, since the inefficiency does not decrease monotonically, the inefficiency at $a\approx1.9\lambda$ is smaller than at the preceding minimum around $a\approx\sqrt{2}\lambda$, demonstrating that sometimes increasing the lattice spacing can reduce the inefficiency. Figure \ref{realistic}b shows that a cavity with finesse $\mathcal{F}=10000$ and a beam waist $w/\lambda=5$ reduces the inefficiency to $\epsilon\sim3.5\times 10^{-4}$.

\emph{Discussion.---}
In this work, we analyzed the problem of coupling atomic tweezer arrays to light via an optical cavity, taking into account collective scattering. We showed that the coupling can be enhanced by the cavity finesse, and discussed its optimization by a proper choice of the lattice spacing near the diffraction resonances. These results can be generalized to array structures beyond that of a square lattice, such as a triangular lattice, for which we find similar results. Our findings are essential for using the array as an efficient light-matter interface in various quantum tasks, such as for interconnecting multiple arrays through flying photons. Furthermore, it opens the door for quantum optics applications with tweezer arrays, such as quantum non-linear optics with  Rydberg atoms or light storage. 

\begin{acknowledgments}
We acknowledge financial support from the Israel Science Foundation (ISF), 
the Directorate for Defense Research and Development (DDR\&D),
the US-Israel Binational Science Foundation (BSF) and US National Science Foundation (NSF), 
the Minerva Foundation with funding from the Federal German Ministry for Education and Research,
the Center for New Scientists at the Weizmann Institute of Science, the Council for Higher Education (Israel), and QUANTERA (PACE-IN), the Estate of Louise Yasgour.  This research is made possible in part by the historic generosity of the Harold Perlman Family.
\end{acknowledgments}

\newpage
\appendix

\end{document}